\documentclass[11pt,a4paper]{article}

\usepackage{amsmath,amssymb,url,cite,ifpdf,slashed,multirow}

\ifpdf                                     
  \usepackage{graphicx}                    
\else                                      
  \usepackage[dvipdfmx]{graphicx}          
\fi

\newcommand{\TeV}{\,{\rm TeV}}
\newcommand{\GeV}{\,{\rm GeV}}

\newcommand{\invfb}{\,{\rm fb^{-1}}}

\newcommand{\s}[1]{_{\rm #1}}
\newcommand{\mL}{m_{\rm L}}
\newcommand{\mR}{m_{\rm R}}

\newcommand{\Order}{\mathop{\mathcal{O}}}
\newcommand{\PT}{p\s T}
\newcommand{\vc}[1]{\boldsymbol{#1}}

\newcommand{\MET}{\slashed{E}\s T}
\newcommand{\vMET}{\slashed{\vc{E}}\s T}

\makeatletter
\newcommand{\@authornote}[2]{{\def\thefootnote{\fnsymbol{footnote}}\setcounter{footnote}{#1}#2\setcounter{footnote}{0}}}
\newcommand{\authornotemark}[1]{\@authornote#1{\addtocounter{footnote}{-1}\footnotemark}}
\newcommand{\authornotetext}[2]{\@authornote#1{\footnotetext{#2}}}
\makeatother

\begin{document}

\begin{titlepage}

\begin{flushright}
UT--13--10
\end{flushright}

\vskip 3cm
\begin{center}
{\Large \bf
Muon $g-2$ vs LHC
in Supersymmetric Models
}
\vskip .75in

{\large
Motoi Endo,
Koichi Hamaguchi,\\
Sho Iwamoto\authornotemark{1},
Takahiro Yoshinaga
}
\vskip 0.25in
\authornotetext{1}{Research Fellow of the Japan Society for the Promotion of Science}

{\it Department of Physics, The University of Tokyo,
   Tokyo 113--0033, Japan
}

\end{center}
\vskip .5in

\begin{abstract}
There is more than $3\sigma$ deviation between the experimental and theoretical results of the muon $g-2$. 
This suggests that some of the SUSY particles have a mass of order 100 GeV. 
We study searches for those particles at the LHC with particular attention to the muon $g-2$. 
In particular, the recent results on the searches for the non-colored SUSY particles are investigated in the parameter region where the muon $g-2$ is explained. 
The analysis is independent of details of the SUSY models. 
Future prospects of the collider searches are also discussed.
\end{abstract}

\end{titlepage}
\setcounter{page}{2}

\section{Introduction}
\label{sec:introduction}

The ATLAS and the CMS collaborations have reported observations of a new particle
with a mass of about 126 GeV, which is considered to be the Standard Model (SM) Higgs boson~\cite{aad:2012gk,Chatrchyan:2012gu}.
If the particle is indeed the Higgs boson, the SM inevitably involves the hierarchy problem.
This unnaturalness indicates that there lies physics beyond the SM.

There is another indication for physics beyond the SM.
The precise measurement of the muon anomalous magnetic moment (the muon $g-2$)~\cite{g-2_bnl2010} has shown discrepancy from the SM prediction~\cite{g-2_hagiwara2011,g-2_davier2010}.
With dedicated efforts to determine hadronic contributions precisely, the latest result is 
\begin{equation}
 \Delta a_\mu \equiv a_\mu({\rm exp}) - a_\mu({\rm SM})= (26.1 \pm 8.0) \times 10^{-10},
 \label{eq:g-2_deviation}
\end{equation}
where the hadronic vacuum polarization is quoted from Ref.~\cite{g-2_hagiwara2011}, and the hadronic light-by-light contribution is from Ref.~\cite{Prades:2009tw}. 
The muon $g-2$ anomaly indicates physics beyond the SM at more than $3\sigma$ level. 
Moreover, the difference is as large as the SM electroweak contribution, $a_\mu({\rm EW}) = (15.4 \pm 0.2) \times 10^{-10}$~\cite{Czarnecki:2002nt}.
If new physics is responsible for the discrepancy, its contribution is naively estimated as $\delta a_\mu\sim(\alpha\s{NP}/4\pi)\times(m_\mu^2/m\s{NP}^2)$, where $\alpha_{\rm NP}$ is a coupling constant of new particles to the muon, and $m_{\rm NP}$ is a typical scale of their masses.
Thus, new physics around the TeV scale is required to involve strong couplings with the muon in order to solve the muon $g-2$ anomaly.

The supersymmetry (SUSY) is a promising candidate for the TeV-scale new physics.
The minimal SUSY Standard Model (MSSM) not only solves the hierarchy problem but also yields sizable contributions to the muon $g-2$~\cite{lopez:1993vi,chattopadhyay:1995ae,moroi:1995yh}.
The SUSY contributions to the muon $g-2$ are naturally enhanced by $\tan\beta \equiv \langle H_u \rangle/\langle H_d \rangle$. As we will see in Sec.~\ref{sec:muon_g-2}, the muon $g-2$ anomaly is solved if the superparticles (smuons, neutralinos, and charginos) are around $\Order(100)\GeV$ for $\tan\beta = \Order(10)$.

Superparticles are searched for at the LHC.
Since none of them has been discovered, colored superparticles are considered to be heavier than $\Order(1)\TeV$~\cite{ATLAS2012109,recentcms}.
Moreover, the Higgs boson mass of $126\GeV$ indicates the scalar tops to be as heavy as $\Order(1\text{--}10)\TeV$~\cite{heavy_stops} and/or to have a large trilinear coupling to the up-type Higgs boson~\cite{Okada:1990gg}.
These two results naively contradict with the indication of the muon $g-2$ anomaly that superparticles have a mass of $\Order(100)\GeV$. 
In fact, some of the representative SUSY-breaking mediation mechanisms such as the minimal supergravity and the gauge mediation cannot  satisfy the above three phenomenological requirements simultaneously, (i) providing the Higgs boson mass of $126\GeV$, (ii) avoiding the constraints from the direct searches at the LHC, and (iii) solving the muon $g-2$ anomaly (see Ref.~\cite{Endo:2011gy} for a study on the minimal supergravity).
From the viewpoint of model building, the muon $g-2$ is reconciled with the Higgs boson mass of $126\GeV$ by extending the MSSM so that extra contributions to the Higgs potential appear~\cite{Endo:2011gy,Endo2012vectorlike,EIY}.

The above inconsistency implies a split mass spectrum of the superparticles as
\begin{equation}
m_{\tilde{q}} \gg m_{\tilde{\ell}},\, m_{\tilde{\chi}^\pm},\, m_{\tilde{\chi}^0},
\label{eq:rough_spectrum}
\end{equation}
where the squarks are favored to be relatively heavy to explain the Higgs boson mass and to satisfy the LHC bounds, while the sleptons, neutralinos and charginos are to be light to explain the muon $g-2$ discrepancy.
This scenario forces us to change strategy of the SUSY search.
The searches based on productions of colored superparticles, which have been the standard, are not always promising, because the particles can be away from the LHC reach without diminishing a virtue of the SUSY as a solution to the muon $g-2$ anomaly.
In this letter, we investigate the LHC signatures assuming the mass hierarchy in Eq.~(\ref{eq:rough_spectrum}),
with particular attention to the muon $g-2$.
We examine the direct productions of the superparticles which are relevant for the muon $g-2$.~\footnote{Those superparticles can also be probed indirectly via the precision measurements~\cite{Cho:2011rk}.}
The ATLAS and CMS collaborations respectively reported results of their searches for this production channel~\cite{ATLAS2012154,CMSPASSUS12022}. We will apply the result by the ATLAS to the above scenario, and will see that these searches are particularly important in the parameter regions where the muon $g-2$ is explained.

In Sec.~\ref{sec:muon_g-2}, the SUSY contributions to the muon $g-2$ is reviewed, and the parameter regions in which the discrepancy \eqref{eq:g-2_deviation} is explained are clarified.
The detailed mass spectra and parameter spaces are introduced in Sec.~\ref{sec:mass}, and relevant SUSY searches at the LHC are summarized in Sec.~\ref{sec:LHC}.
In Sec.~\ref{sec:result}, current LHC bounds on the muon $g-2$ parameter spaces are shown, and future prospects are discussed in Sec.~\ref{sec:future}.
The last section is devoted to the conclusion.


\section{Muon $g-2$}
\label{sec:muon_g-2}

The SUSY contributions to the muon $g-2$ are dominated by the chargino--sneutrino and the neutralino--smuon loop diagrams. At the leading order of $m_W/m_{{\rm{soft}}}$ and $\tan\beta$, where $m_{{\rm{soft}}}$ represents SUSY-breaking masses and the Higgsino mass $\mu$, they are evaluated as~\cite{moroi:1995yh}
\begin{align}
 \Delta a_{\mu }(\tilde{W}, \tilde{H}, \tilde{\nu}_\mu)
 &= \frac{\alpha_2}{4\pi} \frac{m_\mu^2}{M_2 \mu} \tan\beta\cdot
f_C
 \left( \frac{M_2 ^2}{m_{\tilde{\nu }}^2}, \frac{\mu ^2}{m_{\tilde{\nu }}^2}  \right) , 
 \label{eq:WHsnu} \\
 \Delta a_{\mu }(\tilde{W}, \tilde{H},  \tilde{\mu}_L) 
 &= - \frac{\alpha_2}{8\pi} \frac{m_\mu^2}{M_2 \mu} \tan\beta\cdot
 f_N
 \left( \frac{M_2 ^2}{m_{\tilde{\mu }_L}^2}, \frac{\mu ^2}{m_{\tilde{\mu }_L}^2} \right),  
 \label{eq:WHmuL}  \\ 
 \Delta a_{\mu }(\tilde{B},\tilde{H},  \tilde{\mu }_L) 
  &= \frac{\alpha_Y}{8\pi} \frac{m_\mu^2}{M_1 \mu} \tan\beta\cdot
 f_N 
 \left( \frac{M_1 ^2}{m_{\tilde{\mu }_L}}, \frac{\mu ^2}{m_{\tilde{\mu }_L}} \right),   
 \label{eq:BHmuL} \\ 
  \Delta a_{\mu }(\tilde{B}, \tilde{H},  \tilde{\mu }_R) 
  &= - \frac{\alpha_Y}{4\pi} \frac{m_{\mu }^2}{M_1 \mu} \tan \beta \cdot
  f_N \left( \frac{M_1 ^2}{m_{\tilde{\mu }_R}^2}, \frac{\mu ^2}{m_{\tilde{\mu }_R}^2} \right), \label{eq:BHmuR} \\
 \Delta a_{\mu }(\tilde{\mu }_L, \tilde{\mu }_R,\tilde{B}) 
 &= \frac{\alpha_Y}{4\pi} \frac{m_{\mu }^2 M_1 \mu}{m_{\tilde{\mu }_L}^2 m_{\tilde{\mu }_R}^2}  \tan \beta\cdot
 f_N \left( \frac{m_{\tilde{\mu }_L}^2}{M_1^2}, \frac{m_{\tilde{\mu }_R}^2}{M_1^2}\right), \label{eq:BmuLR} 
\end{align}
where $m_\mu$ is the muon mass, while $\alpha_Y$ and $\alpha_2$ are the fine structure constants of the SM U(1)$_Y$ and the SU(2)$_L$ gauge symmetries, respectively. The loop functions are defined as\footnote{
The functions, $f_C$ and $f_N$, are reduced from the functions, $J_5$ and $I_4$, in Ref.~\cite{moroi:1995yh}.
}
\begin{align}
&f_C(x,y)= xy
\left[
\frac{5-3(x+y)+xy}{(x-1)^2(y-1)^2}
-\frac{2\log x}{(x-y)(x-1)^3}
+\frac{2\log y}{(x-y)(y-1)^3}
\right]\,,
\\
&f_N(x,y)= xy
\left[
\frac{-3+x+y+xy}{(x-1)^2(y-1)^2}
+\frac{2x\log x}{(x-y)(x-1)^3}
-\frac{2y\log y}{(x-y)(y-1)^3}
\right]\,.
\end{align}
They satisfy $0\le f_{C,N}(x,y) \le 1$ and are monochromatically increasing for $x>0$ and $y>0$.
In the limit of degenerate masses, they satisfy $f_C(1,1)=1/2$ and $f_N(1,1)=1/6$.
The arguments in the left-hand side of Eqs.~\eqref{eq:WHsnu}--(\ref{eq:BmuLR}) show the superparticles which propagate in each loop diagram.
If one of them decouples, the corresponding SUSY contribution is suppressed.
Eq.~(\ref{eq:WHsnu}) comes from the chargino--sneutrino diagrams, and Eqs.~(\ref{eq:WHmuL})--(\ref{eq:BmuLR}) are the neutralino--smuon contributions. 
We can easily see that the parameters
\begin{align}
M_1, M_2, \mu, m_{\tilde{\mu }_L}, m_{\tilde{\mu }_R}, \tan \beta,
\label{eq:parameters}
\end{align}
are relevant, where $M_{1,2}$ are the bino and wino masses, and $m_{\tilde{\mu }_{L,R}}$ are masses of the left- and right-handed smuons.
Note that the sneutrino mass $m_{\tilde{\nu }}$ is related to $m_{\tilde{\mu}_L}$ by the SU(2)$_L$ symmetry as $m^2_{\tilde{\nu }} =  m^2_{\tilde{\mu}_L} + m^2_W \cos 2\beta $.
Numerically, the SUSY contributions are evaluated as
\begin{align}
 \Delta a_{\mu }(\tilde{W},\tilde{H}, \tilde{\nu}_{\mu} ) 
&\simeq 
\phantom{+}15\phantom{.0} \times 10^{-9}
\left(\frac{\tan\beta}{10}\right)
\left(\frac{(100\GeV)^2}{M_2\mu}\right)
\left(\frac{f_C}{1/2}\right),  \label{eq:WHLnu_N}  \\
 \Delta a_{\mu }(\tilde{W},\tilde{H}, \tilde{\mu }_L) 
&\simeq 
-2.5 \phantom{0} \times 10^{-9}
\left(\frac{\tan\beta}{10}\right)
\left(\frac{(100\GeV)^2}{M_2\mu}\right)
\left(\frac{f_N}{1/6}\right),  \label{eq:WHL_N}  \\
 \Delta a_{\mu }(\tilde{B},  \tilde{H},  \tilde{\mu }_L) 
  &\simeq
\phantom{+}0.76\times 10^{-9}
\left(\frac{\tan\beta}{10}\right)
\left(\frac{(100\GeV)^2}{M_1\mu}\right)
\left(\frac{f_N}{1/6}\right), \label{eq:BHL_N} \\
 \Delta a_{\mu }(\tilde{B}, \tilde{H},  \tilde{\mu }_R) 
 &\simeq
-1.5\phantom{0}\times 10^{-9}
\left(\frac{\tan\beta}{10}\right)
\left(\frac{(100\GeV)^2}{M_1\mu}\right)
\left(\frac{f_N}{1/6}\right), \label{eq:BHR_N} \\
 \Delta a_{\mu }(\tilde{\mu }_L, \tilde{\mu }_R,\tilde{B}) 
 &\simeq
\phantom{+}1.5\phantom{0}\times 10^{-9}
\left(\frac{\tan\beta}{10}\right)
 \left(\frac{(100\GeV)^2}{m_{\tilde{\mu }_L}^2 m_{\tilde{\mu }_R}^2/M_1\mu }\right)
\left(\frac{f_N}{1/6}\right). \label{eq:BLR_N}
\end{align}

The SUSY contributions to the muon $g-2$ are enhanced when $\tan\beta$ is large and $m_{{\rm{soft}}}$ is small.
For $m_{{\rm{soft}}} = \Order(100)\GeV$ and $\tan \beta = \Order(10)$, they become $\Order(10^{-9})$, which can explain Eq.~\eqref{eq:g-2_deviation}.
Since the experimental value is larger than the SM prediction, $\mathop{\rm sgn}(M_{1,2}\,\mu) > 0$ is favored in most of the parameter region. 
Throughout this letter, the gaugino masses are taken positive.

Let us investigate the SUSY contributions in detail to specify the parameter regions which should be searched for at the LHC.
We will study two representative cases when the muon $g-2$ discrepancy \eqref{eq:g-2_deviation} is explained.
Other possibilities will be discussed in Sec.~\ref{sec:future}. 

The first  case is the one when the chargino--sneutrino contribution (\ref{eq:WHLnu_N}) dominates the SUSY contributions.
As can be seen in Eqs.~(\ref{eq:WHLnu_N})--(\ref{eq:BLR_N}),
this is valid when the relevant mass parameters of Eq.~\eqref{eq:parameters} are nearly degenerate.
In fact, in wide SUSY models which explain the anomaly with non-decoupling Higgsinos, the chargino--sneutrino contribution (\ref{eq:WHLnu_N}) dominates.

As the Higgsino mass $\mu$ increases, the chargino--sneutrino contribution decreases, while the neutralino--smuon contribution becomes relevant.
This is because the pure-bino contribution Eq.~\eqref{eq:BmuLR} is enhanced under the presence of a large $\mu$ resulting in a large left--right mixing in the smuon mass matrix.
The other contributions in Eqs.~(\ref{eq:WHsnu})--(\ref{eq:BHmuR}) are suppressed, where the Higgsino propagates in the diagrams.
In this situation, which we will take as the second  case, the muon $g-2$ discrepancy \eqref{eq:g-2_deviation} is explained when both $m_{\tilde{\mu }_L}$ and $m_{\tilde{\mu }_R}$ are $\Order(100)\GeV$ as will be seen in the next section.

In this letter, we will study the SUSY searches at the LHC for the parameter regions of these two cases.
The soft SUSY-breaking parameters are set at the SUSY scale, and we will not specify the models to realize them.
It is interesting to mention that such parameter regions are realized in wide SUSY models which solve the muon $g-2$ anomaly~(see e.g.~\cite{Endo:2011gy,Endo2012vectorlike,EIY}), in which much ingenuity is exercised in enhancing the Higgs boson mass.
As we take model independent approach, the following LHC analysis will be independent of the mechanisms which realize the Higgs boson mass of $126\GeV$.


\section{Mass spectrum}
\label{sec:mass}

Let us specify the mass spectrum to study LHC phenomena.
We consider two representative cases in light of the muon $g-2$:
(i) the case when the chargino--sneutrino contribution to the muon $g-2$ dominates, and (ii) that with a large $\mu$ parameter where the neutralino--smuon diagram is relevant.
The former contribution is controlled by $(M_2, \mu, m^2_{\tilde{\mu}_L}, \tan\beta)$, and the latter is by $(M_1, \mu, m^2_{\tilde{\mu}_L}, m^2_{\tilde{\mu}_R}, \tan\beta)$, according to Eqs.~\eqref{eq:WHsnu} and \eqref{eq:BmuLR}, respectively.
In particular, the $\mu$ parameter is favored to be relatively small in the former case, while it is fairly large in the latter.

The leading SUSY contributions to the muon $g-2$ are proportional to $\tan\beta$, while the LHC sensitivities in the following sections do not depend much on it.\footnote{
When the staus are light, the branching ratio of the Higgs boson decay into the di-photon can be sensitive to $\tan\beta$ mainly through the left-right mixing of the stau mass matrix.~\cite{Carena:2012xa,Kitahara:2012pb,Carena:2012mw,Kitahara:2013lfa}
\label{footnote:stau}
}
It is fixed as $\tan\beta=40$ throughout the LHC analysis.
As $\tan\beta$ increases, although the SUSY contributions to the muon $g-2$ are enhanced, the bottom and tau Yukawa coupling constants are likely to blow up below the GUT scale.
If $\tan\beta$ is lowered, the superparticle masses must be reduced to solve the muon $g-2$ anomaly.
Then, the LHC searches become easier. 

The soft masses of the left- and right-handed selectrons are respectively set to be the same as those of the smuons, which are denoted by $\mL$ and $\mR$. The results in the following sections are almost independent of the selectron masses.
The third-generation sleptons (staus and tau-sneutrino) 
are irrelevant for the SUSY contributions to the muon $g-2$. They 
are taken to be decoupled in the following analysis to simplify the LHC studies.
Scenarios with light staus will be discussed in Sec.~\ref{sec:future}.

For the gaugino masses, the following two scenarios are analyzed:
(A) they satisfy an approximate GUT relation, $M_1:M_2:M_3 = 1:2:6$, which is 
realized in a wide variety of SUSY models,
and (B) the gluino is decoupled, while the bino and wino masses satisfy the same relation as (A), $2M_1= M_2 \ll M_3$. 
The SUSY contributions to the muon $g-2$ are sensitive to $M_1$ and $M_2$, while $M_3$ barely affects the muon $g-2$ and the Higgs boson mass, but the LHC study depends on it.
In the case (A), superparticles can be searched for by analyzing events with hard-jets plus missing transverse energy, which come from the gluino pair-production, while electroweak productions of superparticles are relevant in the both cases (A) and (B). 
We will study both of (A) and (B) separately for the cases, (i) and (ii).
The bino and wino masses are set to satisfy $2M_1= M_2$ throughout the analysis, and general cases will be discussed in Sec.~\ref{sec:future}.

It is assumed that the squarks are decoupled from the LHC sensitivities in the analysis.
We set their soft masses as $(m^2)_Q=(m^2)_{\bar U}=(m^2)_{\bar D}=(7\TeV)^2$.
They satisfy the current bounds on the superparticle masses obtained at the LHC~\cite{ATLAS2012109,recentcms}, and also, heavy stops are preferred to realize the Higgs boson mass of $126\GeV$~\cite{heavy_stops}. 
The following results do not change as long as the masses are large enough.
Note that lighter squarks just tighten the LHC constraints, because their productions contribute to the SUSY events. 

The remaining SUSY parameters are the CP-odd Higgs mass, $m_A$, and scalar trilinear couplings ($A$-terms). 
They are relevant neither for the muon $g-2$ nor the SUSY searches at the LHC. 
In the analysis, they are fixed as $m_A=1.5\TeV$ and $A_{t,b,\tau}=0$, as a reference.

To summarize, the SUSY signals will be studied at the LHC in the following parameters.
The parameters relevant  to the muon $g-2$ and the LHC signatures are
\begin{align*}
&
[m^2]_{L_1}=[m^2]_{L_2} \equiv \mL,~~~
[m^2]_{\bar E_1}=[m^2]_{\bar E_2} \equiv \mR,~~~
\tan\beta=40,\\
&
\text{(A)}~M_1:M_2:M_3 = 1:2:6,~~~{\rm or}~~~
\text{(B)}~2M_1= M_2 \ll M_3,
\end{align*}
and the other parameters are set as
\begin{align*}
&
[m^2]_{Q}=[m^2]_{\bar U}=[m^2]_{\bar D}=(7\TeV)^2,~~~
[m^2]_{L_3}=[m^2]_{\bar E_3} =(3\TeV)^2,\\
&
m_A=1.5\TeV, \qquad A_t=A_b=A_\tau=0.
\end{align*}
Consequently, there are four free parameters left: ($M_2$, $\mu$, $\mL$, $\mR$).
In order to search for the two scenarios, (i) the chargino--sneutrino dominance and (ii) the neutralino--smuon dominance with large $\mu$, we will take $(M_2,\mL)$ to be free, and the other two parameters are chosen as
\begin{equation}
 (\mu,\mR)=
\{(M_2,3\TeV), (2M_2,3\TeV), (0.5M_2,3\TeV), (2\TeV,1.5\mL)\}.
\label{eq:mumr}
\end{equation}
In the first three sets where $\mR = 3\TeV$, the contribution (i) dominates over the SUSY contributions to the muon $g-2$. 
The last set corresponds to the case (ii).

In the above parameter space, the lightest superparticle (LSP) is either the bino-like neutralino, the neutral Higgsino, or the sneutrino.
The LSP is a candidate of the dark matter if the R-parity is conserved. 
In the last set of \eqref{eq:mumr}, $m_R$ is set to be slightly larger than $m_L$ in order to avoid a charged LSP in the whole parameter region, since the charged dark matter is excluded~\cite{Kudo:2001ie}. The sneutrino LSP is also strongly disfavored by the direct searches~\cite{Falk:1994es}.
In the analysis, we will focus on the cosmologically favored regions.
Note that other superparticles such as the gravitino can be lighter than them, and such cases will be discussed in Sec.~\ref{sec:future}.


\section{LHC phenomenology}
\label{sec:LHC}

In order to solve the muon $g-2$ anomaly, the bino and wino masses are as small as $\Order(100)\GeV$.
Then, if we respect the approximate GUT relation, the gluino has a mass less than a few TeV, which is within the reach of the LHC. 
The gluinos are produced at the $pp$ collisions, and decay into neutralinos or charginos with hard jets,
\begin{equation}
pp\to \tilde{g}\tilde{g}\to qq\tilde{\chi}\; qq\tilde{\chi}\to\cdots\,.
\label{eq:gluino-decay}
\end{equation}
Similarly to the minimal supergravity models, searches for hard jets plus missing transverse energy are one of the most promising channels.
Currently, the ATLAS collaboration yields the most stringent bound in this category~\cite{ATLAS2012109} with analyzing their data corresponding to $\int\!\mathcal{L} = 5.8\invfb$ obtained at $\sqrt{s}=8\TeV$.
We interpret this result in this letter, which will be referred to as ``J-search.''

The electroweak gaugino productions (e.g., $pp\to\tilde\chi^{\pm}_1\tilde\chi^0_2$) are also searched for at the LHC.
When sleptons are relatively light, the gauginos can produce hard leptons as 
\begin{equation}
pp\to \tilde{\chi}\tilde{\chi} \to \ell\tilde{\ell}\;\ell\tilde{\ell}
 \to \ell \ell\tilde{\chi}\;\ell \ell\tilde{\chi}\,.
\end{equation}
As the muon $g-2$ prefers such lighter sleptons, the multi-lepton signatures are another promising channel.
The ATLAS collaboration recently reported searches for events with three leptons plus missing transverse energy in the data of $\int\!\mathcal{L} = 13.0\invfb$ at $\sqrt{s}=8\TeV$~\cite{ATLAS2012154}.
We interpret the result, which will be referred to as ``L-search.''
The multi-lepton signatures are particularly important, because they are yielded by the superparticles relevant for the muon $g-2$, i.e., sleptons, charginos, and neutralinos, and appear even in the decoupled gluino scenario (B).

Those two ATLAS results are interpreted in the analysis with Monte Carlo simulation.
On each model point, the mass spectra generated with {\tt SOFTSUSY\,3.4}~\cite{SOFTSUSY} are passed to {\tt SUSY-HIT\,1.3}~\cite{SUSYHIT} to calculate decay tables of the superparticles.
Here, the soft SUSY-breaking parameters are set at the SUSY scale, $M\s{SUSY}=(m_{\tilde t_1}m_{\tilde t_2})^{1/2}$.
The kinematical distributions of SUSY events are simulated by {\tt Pythia\,6.4}~\cite{Pythia6.4} with the ATLAS {\tt MC09} tune~\cite{ATLASPHYSPUB2010002}.
The parton distribution functions (PDFs) are obtained from the {\tt CTEQ6L1} set~\cite{PDFCTEQ6}.
As the squarks are assumed to be decoupled, superparticles are produced via the gluino pair-production ($pp\to\tilde g\tilde g$) and the electroweak productions of neutralinos, charginos and sleptons ($pp\to\tilde\chi\tilde\chi,\tilde\ell\tilde\ell^*$).
For the gluino productions, the cross sections are normalized by NLO $K$-factors obtained with {\tt Prospino\,2}~\cite{Prospinoweb,ProspinoSG}, where the {\tt CTEQ6L1} and {\tt CTEQ6.6M} PDFs~\cite{PDFCTEQ6} are used.
For the electroweak channels, the normalization factor is set as $K=1.2$, which is a typical value in the parameter space studied in the next section.
In addition, the SUSY contribution to the muon $g-2$ is  evaluated by {\tt FeynHiggs}~\cite{FeynHiggs}, which includes two-loop contributions.

Detector simulation is employed with {\tt Delphes\,2.0}~\cite{Delphes}.
We assume trigger efficiencies to be $100\%$ for simplicity.
As for calorimeter configuration and resolutions, {\tt Delphes} original parameter card for the ATLAS detector is used.

Following the ATLAS analysis, jet reconstruction is done with the anti-$k_t$ jet clustering algorithm~\cite{anti-kt} with the distance parameter $R=0.4$. 
{\tt FASTJET}~\cite{FASTJET} is utilized.
We also simulate $b$-tagging, used as $b$-veto in the L-search, with simplified efficiency and fake rates estimated from Refs.~\cite{ATLAS2012040,ATLAS2012043}.
The missing transverse energy is read from {\tt Delphes} output without modification.

We take lepton detection efficiencies into account, which are important particularly for the L-search.
For electrons, the ATLAS collaboration utilizes three identification criteria.
We consult Ref.~\cite{Aad:2011mk}, and set the efficiency of the ``medium'' criterion as $0.89$ for $\PT>20\GeV$ and $0.75$ for $10\GeV<\PT<20\GeV$, and that of the ``tight'' criterion as $0.74$ for $\PT>20\GeV$ and $0.61$ for $10\GeV<\PT<20\GeV$. $|\eta|$ dependence of the efficiency is ignored.
For muons, we interpret the results in Ref.~\cite{ATLAS2011063}, and set the efficiency (for $\PT>10\GeV$) as $0.82$ for $|\eta|<0.25$ and $0.96$ for $|\eta|>0.25$.
Note that we further impose the lepton isolation criteria in the L-search, and the overlap removal in the J- and L-searches, in the same way as the ATLAS analysis.

The signal regions (SRs) are defined to be the same as those in the original ATLAS analyses.
The details are summarized in the following subsections.
Throughout the analysis, we ignore criteria on the primary vertex, origins of lepton tracks, and jet qualities, which are designed to reduce noise backgrounds.
The $CL\s{s}$ method is used to derive exclusions for each model point.
The numbers of the signal events in SRs are compared to the corresponding upper bounds obtained in the ATLAS reports.

The analysis procedures are validated by comparing the simulations with the ATLAS results.
We checked that calculated effective mass distributions of the signal events agree with those of the ATLAS in the J-search, and those of the missing transverse energy are reproduced for the L-search.
Exclusion plots are also simulated by using the upper bounds on the signal events obtained in the ATLAS analysis, 
and become consistent with those in the ATLAS papers.

\subsection{J-SEARCH}
 \begin{table}[tp]
 \begin{center}
   \caption{Definition of SRs of the J-search.
Among the original 12 SRs~\cite{ATLAS2012109}, the SRs which are relevant for our exclusions are shown.
The effective mass $m\s{eff}^{(n)}$ ($m\s{eff}^{\rm inc}$) is defined as a scalar sum of $\MET$ and $\PT$'s of the leading $n$-jets (all jets with $\PT>40\GeV$).
In the last two rows of C and E, the three values mean different SRs, called as `tight', `medium', and `loose', respectively.
}
  \label{tab:2012-109}\def\arraystretch{1.3}
 \begin{tabular}[t]{|c|c|c|c|}\hline
       & \multicolumn{3}{|c|}{Signal Regions} \\\cline{2-4}
   & C & D & E \\
       & ($\ge4$-jets) & ($\ge5$-jets) & ($\ge6$-jets)\\\hline
 $\#$ leptons & \multicolumn{1}{|c}{}&\multicolumn{1}{c}{$=0$}&\multicolumn{1}{c|}{} \\\hline
 $\MET$ [GeV] $>$ & \multicolumn{1}{|c}{}&\multicolumn{1}{c}{160}&\multicolumn{1}{c|}{} \\\hline
 $\PT(j_1)$ [GeV]~~$>$ & \multicolumn{1}{|c}{}&\multicolumn{1}{c}{130}&\multicolumn{1}{c|}{} \\\hline
 $\PT(j_{2,3,4})$ [GeV]~~$>$ & \multicolumn{1}{|c}{}&\multicolumn{1}{c}{60}&\multicolumn{1}{c|}{} \\\hline
 $\PT(j_5)$ [GeV]~~$>$ & --- & 60 & 60 \\\hline
 $\PT(j_6)$ [GeV]~~$>$ & --- & --- & 60 \\\hline
 $\Delta\phi(j_i, \vMET)\s{min}$~~$>$ & \multicolumn{3}{|c|}{$0.6~(i\le 3)$;\quad $0.4$ (other jets with $\PT>40\GeV$)}\\\hline
 \multirow{2}{*}{$\MET/m\s{eff}^{(n)}$~~$>$}&
    0.25\,/\,0.3\,/\,0.3 & 0.15\,/\,---\,/\,---& 0.15\,/\,0.25\,/\,0.3 \\
  & $(n=4)$  & $(n=5)$  & $(n=6)$\\\hline
 $m\s{eff}^{\rm inc}$ [TeV] \,\, $>$&
   1.9\,/\,1.3\,/\,1.0 & 1.7\,/\,---\,/\,--- & 1.4\,/\,1.3\,/\,1.0 \\\hline
  \end{tabular}
 \end{center}
 \end{table}

The J-search is designed to search for pair-productions of the colored superparticles~\cite{ATLAS2012109}.
The event selections are summarized in Table~\ref{tab:2012-109}, where the SRs which are found relevant for the following analysis are shown.
Events are required to have at least 2--6 hard jets with a large missing transverse energy $\MET$.
The efficiencies of triggering, which requires the missing transverse energy of $\ge100\GeV$ or hard jets with $\PT>80\GeV$, are ignored.

\subsection{L-SEARCH}
  \begin{table}[tp]
 \begin{center}
   \caption{Definition of the SRs of the L-search~\cite{ATLAS2012154}.
SR2 targets events of the on-shell $Z$-boson productions from decays of $\tilde\chi_2^0$, while SR1a and SR1b are for $\tilde\chi_2^0$ decaying into leptons (or off-shell $Z$-bosons).
SR1a and SR1b are inclusively defined.
$m\s{SFOS}$ is the invariant mass of a pair of same-flavor opposite-sign leptons (SFOS), and $m\s{T}$ is defined as the transverse mass of the missing energy and the lepton which does not form the SFOS pair which minimizes $|m\s{SFOS}-m_Z|$.
}
  \label{tab:2012-154}\def\arraystretch{1.3}
 \begin{tabular}[t]{|c|c|c|c|}\hline
   & SR1a & SR1b & SR2 \\\hline
  \multirow{2}{*}{$\#$ leptons} & $=3$ & $=3$ & $=3$ \\
              & ($\PT>10\GeV$) & ($\PT>30\GeV$) & ($\PT>10\GeV$) \\\hline
 $\#$ SFOS with $m\s{SFOS}<12\GeV$ & \multicolumn{3}{|c|}{$=0$}\\\hline
 $\#$ SFOS with $m\s{SFOS}>12\GeV$ & \multicolumn{3}{|c|}{$\ge 1$}\\\hline
 $|m\s{SFOS}-m_Z|\s{min}$ & \multicolumn{2}{|c|}{$>10\GeV$} & $<10\GeV$ \\\hline
 $\#$ $b$-jets & \multicolumn{2}{|c|}{0} & any \\\hline
 $\MET$ & \multicolumn{2}{|c|}{$>75\GeV$}& $>120\GeV$ \\\hline
 $m\s T$ & any & $>110\GeV$ & $>110\GeV$ \\\hline
  \end{tabular}
\end{center}
  \end{table}

The L-search is designed to search for electroweak productions of charginos and neutralinos via the s-channel exchange of a virtual SM gauge boson (e.g., $pp\to\tilde\chi^\pm_1\tilde\chi^0_2$). 
The event selections are summarized in Table~\ref{tab:2012-154}.
Events are required to have exactly three hard leptons.
Trigger efficiencies are not taken into account for simplicity.
Note that $b$-jet veto and $m\s T$-cut are necessary to suppress events of SM electroweak backgrounds.


\section{Results}
\label{sec:result}

\begin{figure}[p]
  \begin{center}
   \begin{minipage}[t]{0.45\textwidth}
    \begin{center}
     \includegraphics[width=0.99\textwidth]{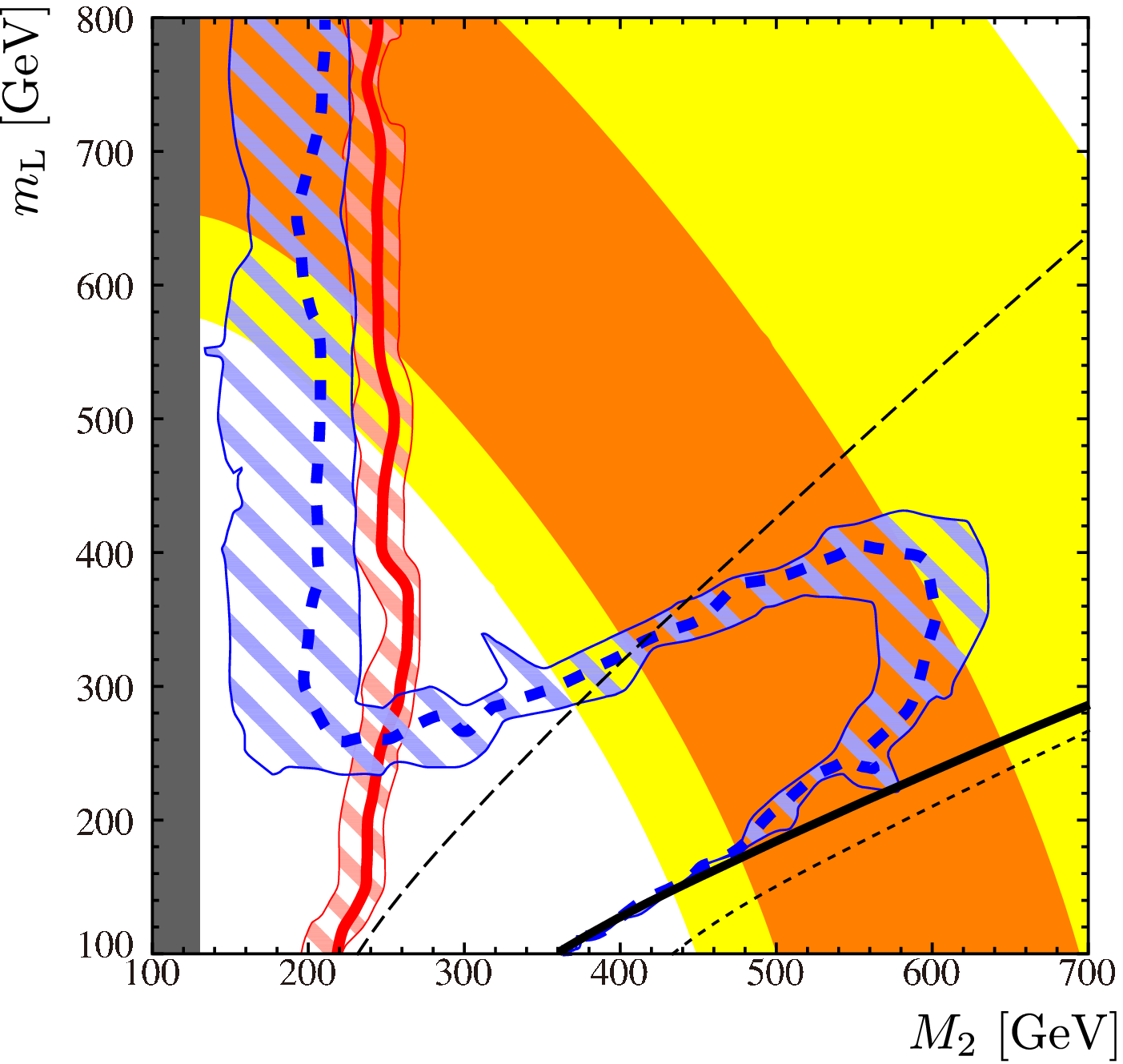}\\(a) $\mu = M_2$, $\mR = 3 \TeV$
    \end{center}   
   \end{minipage}
   \begin{minipage}[t]{0.45\textwidth}
    \begin{center}
     \includegraphics[width=0.99\textwidth]{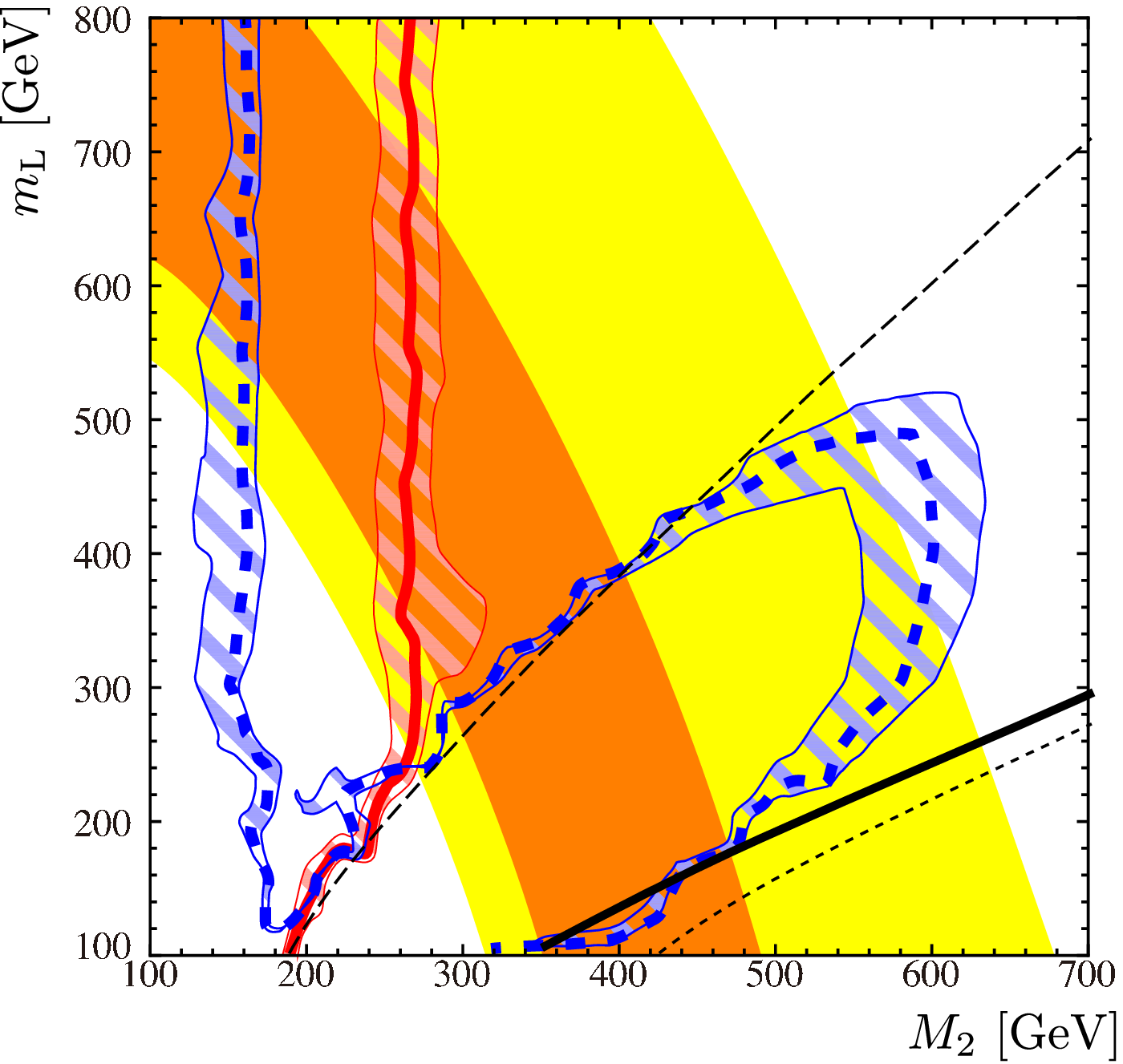}\\(b) $\mu = 2M_2$, $\mR = 3 \TeV$
    \end{center}   
   \end{minipage}
   \\[1em]
   \begin{minipage}[t]{0.45\textwidth}
    \begin{center}
     \includegraphics[width=0.99\textwidth]{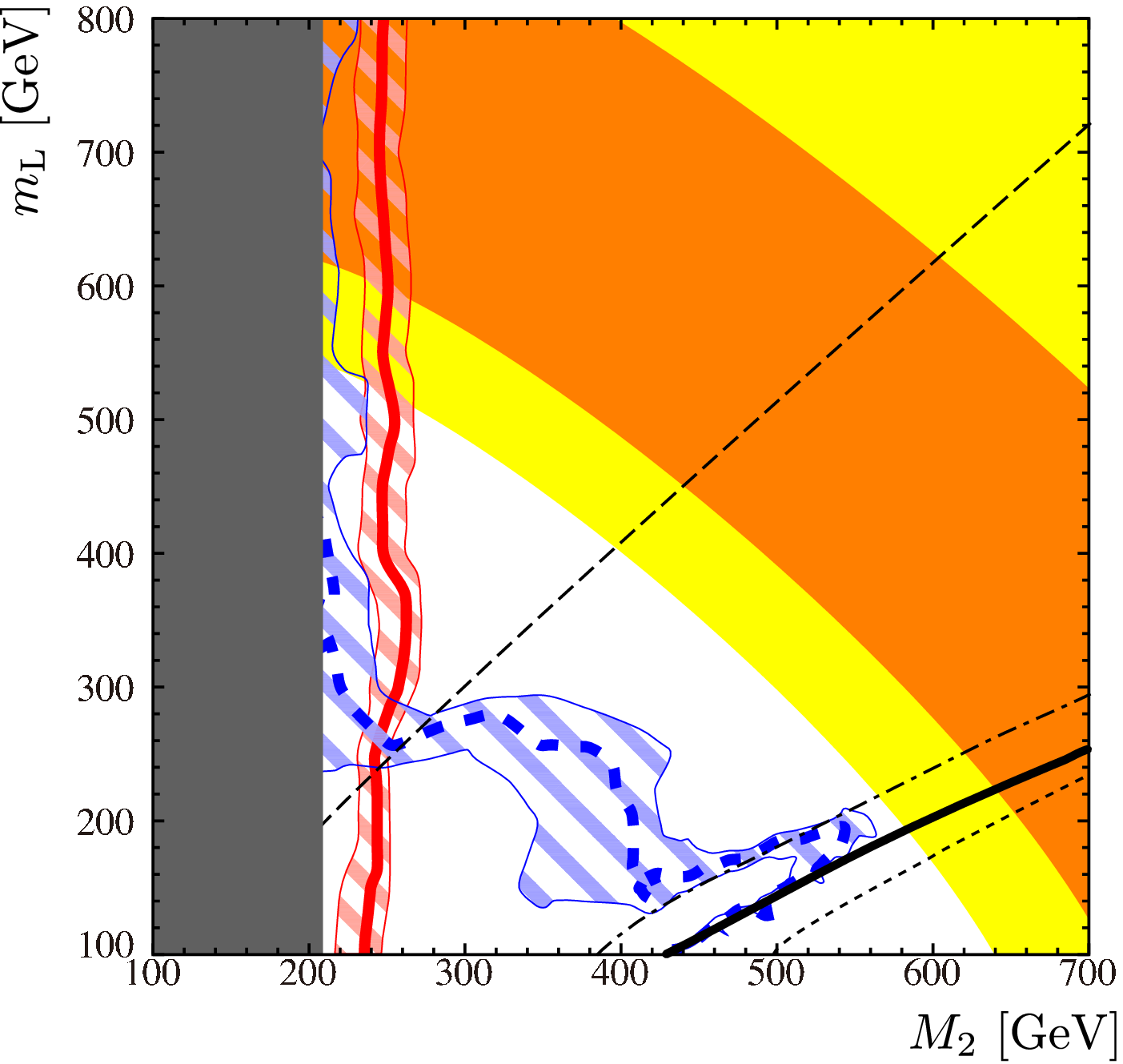}\\(c) $\mu = M_2/2$, $\mR = 3 \TeV$
    \end{center}   
   \end{minipage}
   \begin{minipage}[t]{0.45\textwidth}
    \begin{center}
     \includegraphics[width=0.99\textwidth]{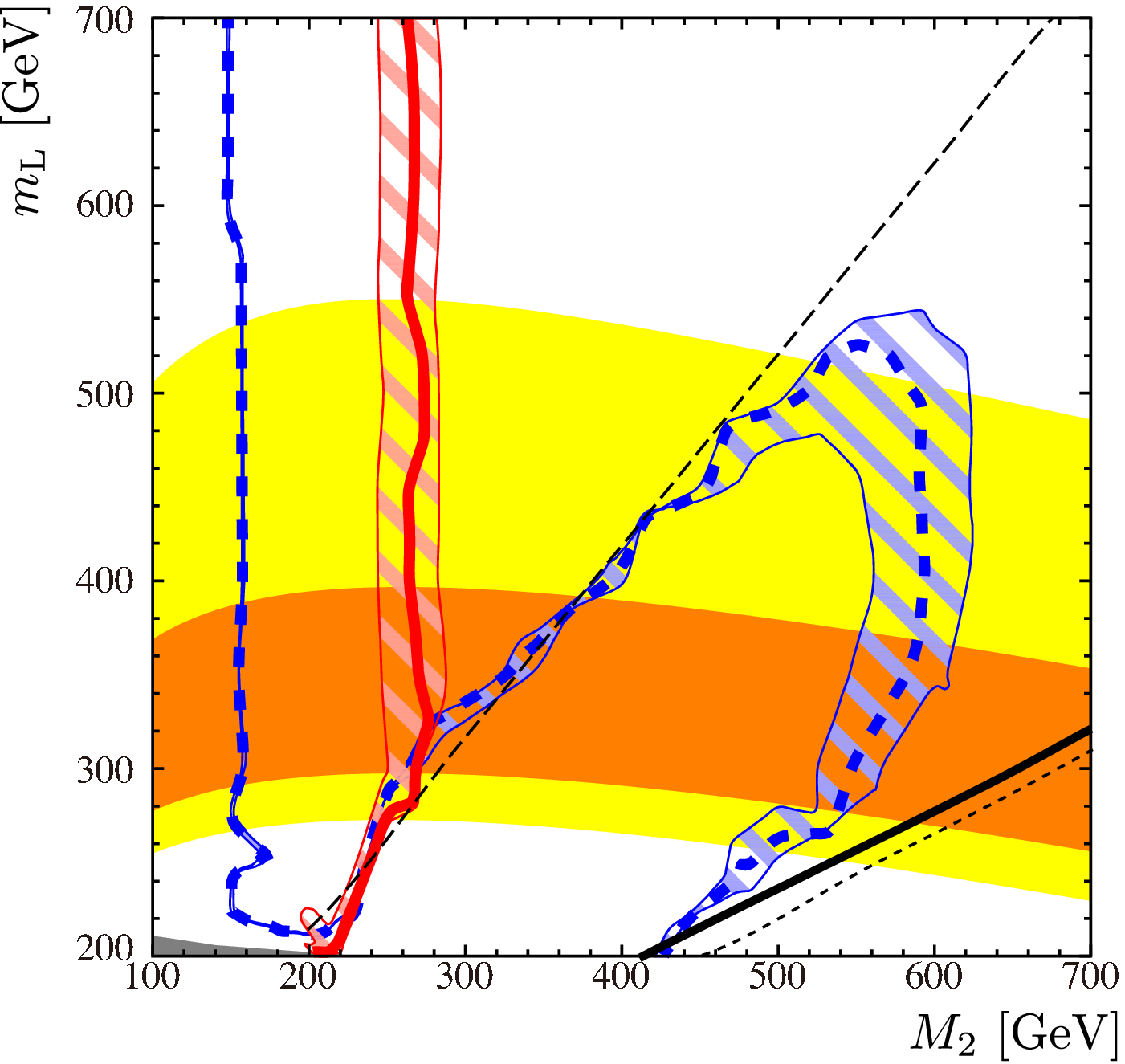}\\(d) $\mu = 2\TeV $, $\mR = 1.5 \mL$
    \end{center}   
   \end{minipage}
  \end{center} 
  \caption{
Current LHC bounds on the SUSY $g-2$ explanations.
The orange (yellow) band shows the region where the SUSY contributions explain the muon $g-2$ discrepancy at the $1\sigma$ ($2\sigma$) level.
The dark gray regions in (a) and (c) are excluded by LEP searches for the neutralinos and charginos.
The regions left to the blue dotted lines are excluded by the L-search.
Assuming the approximate GUT relation for the gaugino masses, the regions left to the red lines are excluded by the J-search.
These exclusions are at $95\%$ CL, and the theoretical uncertainty of $\pm30\%$ is included in the hatched regions.
The LSP is the lightest neutralino in the regions above the black thick lines, while the sneutrino is lightest below them.
Sleptons become lighter than neutralinos below the black dashed or dotted lines (see the text for details).
}
  \label{fig:result}
\end{figure}

The LHC constraints and the muon $g-2$ are shown in Fig.~\ref{fig:result}. 
In the orange (yellow) regions, the muon $g-2$ discrepancy \eqref{eq:g-2_deviation} is explained by the SUSY contributions at the $1\sigma$ ($2\sigma$) level. 
The chargino--sneutrino contribution dominates in the figures (a), (b) and (c),
where the parameters are set as 
$(\mu,\mR) = (M_2,3\TeV), (2M_2,3\TeV)$ and $(0.5M_2,3\TeV)$, respectively.
It is found that the contribution is suppressed as the wino mass increases. 
On the other hand, the neutralino--smuon contribution dominates in the figure (d) with $(\mu,\mR) = (2\TeV,1.5\mL)$. 
The muon $g-2$ is explained even with large $M_2$, while the SUSY contribution decreases as the left- and right-handed slepton masses increase.

The LHC exclusion limits are displayed in Fig.~\ref{fig:result} with the red-solid and blue-dotted lines at the 95\% confidence level (CL).
Including 30\% errors of PDF and scale uncertainties, the limits distribute in the red/blue hatched regions.
Assuming the approximate GUT relation, the regions left to the red lines are excluded by the J-search, and those to the blue lines are constrained by the L-search.
If the gluino is decoupled, the red lines disappear and only the L-search provides a constraint, because the J-search targets the gluino-pair productions.
The bounds from the L-search with the approximate GUT relation are almost identical to those in the gluino decoupling limit. 
It is found that the SUSY searches at the LHC start to constrain the muon $g-2$ (orange or yellow) regions. 

The muon $g-2$ regions are also constrained by  cosmology. 
Noting that the bino mass is set to be a half of the wino mass, the sneutrino becomes the LSP below the black solid lines in Fig.~\ref{fig:result}.\footnote{The slepton pole masses become  larger than the soft mass $\mL$ due to radiative corrections, which are sizable since the soft SUSY-breaking parameters are set at $M\s{SUSY}=(m_{\tilde t_1}m_{\tilde t_2})^{1/2} \simeq 7\TeV$. }
Such parameter regions are disfavored by the dark matter searches~\cite{Falk:1994es}. 
On the other hand, the LSP in the regions above the lines is the lightest neutralino. 
They are constrained by the J- and L-searches at the LHC.

The LHC constraints are understood as follows. 
The J-search targets the gluino-pair productions. 
Among the 12 SRs defined in Ref.~\cite{ATLAS2012109}, those with higher multiplicity of jets are relevant for the exclusion (see Table.~\ref{tab:2012-109}).
Some of the jets originate in the gluino decay into charginos and neutralinos \eqref{eq:gluino-decay}.
Others come from decays of the SM bosons that are decay products of the charginos and neutralinos.
Also, the processes are often associated with initial/final state radiations. 
Assuming the approximate GUT relation, the gluino mass is related to the wino mass, and the red lines are almost independent of the other model parameters in Fig.~\ref{fig:result}. 
The J-search limit $M_2 \gtrsim 250\GeV$ means $M_3 \gtrsim 750\GeV$, leading to $m_{\tilde g} \gtrsim 1.0\TeV$, where $m_{\tilde g}$ is the gluino pole mass.
This is almost the same as the constraint in the minimal supergravity models with heavy squarks~\cite{ATLAS2012109}.
If the ratio $M_3/M_2$ is varied, the exclusion lines shift correspondingly. 

The L-search targets the electroweak superparticle productions.
In particular, the wino productions are significant in the parameter regions in Fig.~\ref{fig:result}. 
The search works only when hard leptons are produced at the wino decay. 
In the regions below the black dashed lines, where sleptons are lighter than the wino-like chargino/neutralino, the wino can decay into an on-shell slepton with a lepton, and the slepton produces another lepton when it decays. 
Then, the SR1b provides a tight constraint. 
Above the lines, the wino mainly decays into the SM bosons as long as the channels are open.
In such regions, no limit is obtained by the L-search.
The search becomes effective again if the mass splitting between the bino and the wino is small, where the wino decays into three-body final states, including leptons. 
Since $M_1$ is set to be $M_2/2$, the limit is determined by the wino mass,
as can be seen from the vertical blue lines in Fig.~\ref{fig:result}. 
Here, the SR1a gives a limit.
On the other hand, 
the search becomes less effective if sleptons are degenerate with the lightest neutralino, because the lepton at the slepton decay becomes soft. 
The charged slepton is degenerate with the lightest neutralino on the black dotted lines.
Note that the regions below the black solid lines are disfavored by cosmology as discussed above.
It is emphasized that the L-search constraints are determined by the superparticles which are relevant for the muon $g-2$.

In Figs.~\ref{fig:result} (a)--(c), the muon $g-2$ regions are partly excluded by the J- and/or L-searches. 
Below the black dashed lines, tight constraints are obtained by the latter.
Above them, the limits are determined by the wino mass, i.e., the gluino mass for the J-search and the splitting between $M_1$ and $M_2$ for the L-search.
Comparing the figures to each other, it is found that the muon $g-2$ regions with larger $\mu$ are easily detected at the LHC.
As $\mu$ increases, $M_2$ is required to be smaller to explain the muon $g-2$.
If the sleptons are lighter than the wino, a wide region of the $1\sigma$ parameter space of the muon $g-2$ is already excluded in (a) and (b).
In contrast, almost all the muon $g-2$ regions are allowed by the LHC in (c). 
In particular, the lepton productions are relatively suppressed for $\mu < M_2$ 
even below the black dashed line, because the winos can decay into the Higgsinos, and the Higgsinos rarely produce leptons due to tiny Yukawa couplings. 
The sensitivity is slightly improved below the black dash-dotted line in (c),
where the sleptons are lighter than the second lightest neutralino.
Since $\mu = M_1$, the second lightest neutralino does not decay into the SM bosons, but can produce the sleptons. 
Here, the SR1a provides a limit from the Higgsino direct productions. 

In Figs.~\ref{fig:result} (a)--(c), $\mR$ is set to be $3\TeV$. 
Since the chargino--sneutrino contribution is insensitive to $\mR$, the muon $g-2$ regions do not change so much even if $m_R$ is varied. 
The LHC constraints do not depend much on $\mR$ either.
Even when it is as low as $\mL$, the right-handed selectron and smuon are rarely produced in the cascade decays of the wino and Higgsino, because they are singlet under the SM SU(2)$_L$ symmetry and have tiny Yukawa interactions.
If the mass is too small, the right-handed slepton becomes the LSP, and thus, excluded by the cosmology.

The LHC constraints in Fig.~\ref{fig:result} (d) are similar to those in (b), where the wino composes the second lightest neutralino. 
Above the black dashed line, since the sleptons are limited to be light by the muon $g-2$, a large part of the $1\sigma$ region is already excluded by the J-search, while the L-search bound is determined similarly to (a)--(c).
On the other hand, the muon $g-2$ is explained even with large $M_2$.
Such a parameter region is challenging to be searched for at the LHC. (See also the discussion 
in the next section.)

In Fig.~\ref{fig:result} (d), $\mu$ is set to be $2\TeV$. 
As $\mu$ increases, the muon $g-2$ is enhanced, and thus, larger $M_1$ is allowed. 
However, the upper limit on the slepton masses is not relaxed so much, as noticed from Eq.~\eqref{eq:BmuLR}. 
Moreover, if $\mu$ becomes too large, the electroweak vacuum can be destabilized on the smuon--Higgs plane.
By naively scaling a bound on $\mu$ from the stau-Higgs stability condition~\cite{Hisano:2010re}, the limit is estimated as $|\mu| \lesssim 10\TeV$ for $\mL = \mR = 200\GeV$ and $\tan\beta = 40$.
In contrast, the LHC constraints do not change, because the Higgsinos are decoupled. 
On the other hand, if $\mu$ is set to be smaller, the upper bound on the wino mass gets tightened, and the LHC detection becomes easier.

\section{Future prospects}
\label{sec:future}

The LHC will be upgraded and run at $\sqrt{s} = 13$ or $14\TeV$ with the target luminosity of $\Order(10)\invfb$.
Further upgrades are proposed to realize $\int\!\mathcal{L} = \Order(100)\invfb$ or $\Order(1000)\invfb$.
There are several studies on future sensitivities of the SUSY searches. 
In this section, we discuss future searches for the SUSY parameter regions of the muon $g-2$. 

\begin{description}
 \item[Multi-jet signature]
Multi-jet searches are important for the approximate GUT relation on the gaugino masses.
According to the future sensitivity in Ref.~\cite{Baer:2012vr} based on the minimal supergravity models, the gluino lighter than $1.8\TeV$ and $2.3\TeV$ can be discovered at the $14\TeV$ LHC with $\int\!\mathcal{L} = 300\invfb$ and $3000\invfb$, respectively.
These gluino masses correspond to $M_2 = 480\GeV$ and $630\GeV$ under the approximate GUT relation, which means a large parameter regions in Fig.~\ref{fig:result} can be covered.
Since the sleptons are expected to be light particularly in a large wino mass region, signatures of multi-jets plus leptons may provide a better sensitivity than the expectations based on the minimal supergravity.

 \item[Multi-lepton signature]
Multi-lepton searches works very well for the two cases: when sleptons are lighter than charginos and neutralinos, and when the wino mainly decays into three bodies.
The former case is typically realized in the regions below the black dashed lines with $M_2 \lesssim \mu$ as we saw in Figs.~\ref{fig:result} (a), (b), and (d), where the multi-leptons are provided from the cascade decays of winos.
The latter case corresponds to the left-most regions in (a)--(d), where the three-body decay provides the leptons.
It is important to study future sensitivities of this channel.

 \item[$W+h$ signature from gaugino/Higgsino decays]
If charginos and neutralinos are lighter than sleptons and can decay into the SM bosons, multi-lepton searches are not promising, but rather, we should explore searches for the SM bosons from the gauginos and/or Higgsinos.
For $W+h$ signature from gaugino pair productions, a study within the minimal supergravity framework in Ref.~\cite{Baer:2012vr} shows that the wino masses of $400\GeV$ and $900\GeV$ can be covered at the $14\TeV$ LHC with $\int\!\mathcal{L} = 300\invfb$ and $3000\invfb$, respectively, in one lepton plus multi-bottoms signature.
With this kind of searches, most of the muon $g-2$ regions above the black dashed lines in Fig.~\ref{fig:result} are expected to be examined at the LHC, where the branching ratios of the SM boson channels are similar to those in the minimal supergravity models.

The $W+h$ signature is also expected from higgsino decays as long as Higgsino is heavier than, but not degenerate to, the LSP.
They are produced by the $s$-channel exchange of a virtual electroweak gauge boson, and their production cross sections are comparable to those of the winos.
Thus, the above sensitivity on the wino mass can be naively applied to the Higgsino mass.

 \item[$Z+Z$ or $h+h$ signature]
It may be possible to discover charginos and neutralinos via the $ZZ$ or $hh$ channels with a large missing transverse energy.
Since the bosons can be reconstructed by visible decay products or have sizable branching ratios of the bottom quark productions, the events could be discriminated from the SM backgrounds.
We need further studies to explore such possibilities.

\item[Same-sign leptons]
Very recently, the authors in Ref.~\cite{Baer:2013yha} studied the events with a same-sign lepton pair and no other leptons, obtained from same-sign $W$-bosons. They found that the signature is useful to search for the electroweak superparticle events.
In our setup, this signature appears more frequently because the sleptons are lighter, and thus, may cover the muon $g-2$ regions.

\item[di-lepton channel search]
Even with combining the above techniques, the bottom right corners of Figs.~\ref{fig:result} (c) and (d) are still challenging.
Future multi-lepton searches at the LHC may not be sensitive enough to cover the regions completely since  $M_2$ is very large.
Here, since sleptons are required to be light, di-lepton channels via the pair productions of sleptons might be promising.
The search at the LHC is challenging because  it suffers from huge SM backgrounds and  their production cross sections are much smaller than those of the charginos and the neutralinos in this parameter region.
The ILC is suitable to search for the di-lepton signatures. 

\end{description}

So far, we considered two representative scenarios to explain the muon $g-2$ discrepancy: with dominance of the chargino--sneutrino contribution, and of the pure-bino--smuon diagram.
We have also set that the stau is heavy, and the LSP is within the MSSM. 
Let us touch on other possibilities.

We have imposed $M_1=M_2/2$, but this is not mandatory.
As long as $M_1$ is smaller than, but not degenerate to, $M_2$, and 
the winos are produced at the LHC, 
the above conclusions and discussions hold valid.
On the other hand, if $M_2$ is too large with $M_1$ fixed, the muon $g-2$ regions disappear in Figs.~\ref{fig:result} (a)--(c), whereas those in Fig.~\ref{fig:result} (d) remain.
Searches for di-lepton signatures are necessary as discussed above.
For the case when $M_2<M_1$, the muon $g-2$ regions in Figs.~\ref{fig:result} (a)--(c) does not depend much on $M_1$, while the LHC signatures are changed.
Signatures from the cascade decays of binos are less promising than those of winos due to a smaller cross section of the bino production.
Rather, signatures from the productions of Higgsinos and sleptons are hopeful.
Moreover, a disappearing track at the LHC is expected to be observed. The track comes from the charged wino, which is almost degenerate with the neutral wino and decays into a neutral wino with a soft pion~\cite{ATLAS:2012jp}.

So far, we investigated the parameter regions where the SUSY contribution to the muon $g-2$ is dominated either by the chargino--sneutrino contribution \eqref{eq:WHsnu}  or
 the pure-bino--smuon contribution \eqref{eq:BmuLR}.
Let us mention other possibilities.
First of all, the contribution of Eq.~\eqref{eq:WHmuL} cannot take dominance, since it is always buried in that of Eq.~\eqref{eq:WHsnu}.
Next, if the left-handed smuon is decoupled, only Eq.~\eqref{eq:BHmuR} contributes to the muon $g-2$, where the sign of $M_1\mu$ is favored to be negative.
Lastly, Eq.~\eqref{eq:BHmuL} becomes dominant if the wino and the right-handed smuon are decoupled.
However, the contributions of Eqs.~\eqref{eq:BHmuL} and \eqref{eq:BHmuR}
are quantitatively small as shown in Eqs.~\eqref{eq:BHL_N} and \eqref{eq:BHR_N}, respectively. Thus, the superparticles must be very light, and the scenarios are expected to be examined by searching for the SM boson or di-lepton channels.

The staus and the tau sneutrino were supposed to be decoupled from the LHC phenomenology.
If their masses are comparable to those of the selectron and smuon, physics becomes much involved.
Attention must be paid to the vacuum stability conditions on the stau-Higgs plane especially when $\mu$ is large, i.e., in Fig.~\ref{fig:result} (d)~\cite{Hisano:2010re,Carena:2012mw}.
The decay branching ratio of the Higgs boson into the di-photon may be affected in such parameter spaces (see the footnote \ref{footnote:stau}).
As for the direct searches for the superparticles, since a stau tends to be light due to a large left-right mixing in the stau mass matrix, taus are likely to be produced in the decay chains of the superparticles.
Then the tau reconstruction becomes a central subject, which is one of the challenging topics of the LHC.
Because of the rich phenomenology, these scenarios are worth investigated for future.

Finally let us relax our restriction of the LSP.
We focused on the neutralino LSP to avoid the cosmological constraints. However, they can be circumvented 
if the LSP is a superparticle which is not in the MSSM, such as the gravitino and the axino. 
In those cases, charged particles or sneutrino can be the LSP among the MSSM superparticles (MSSM-LSP), where the LHC signatures change correspondingly.
If a sneutrino is the MSSM-LSP, hard leptons are less produced because the left-handed charged sleptons are degenerate to the sneutrinos. Although sneutrino decays into the LSP are not visible, the pair-production of the sleptons could be found by the mono-jet or mono-photon searches.
In addition, this case might be checked by reconstructing soft particles that are produced by the quasi-degenerate charged slepton when it decays into the sneutrino. This search could be done at the ILC.
On the other hand, when the MSSM-LSP is a charged slepton, it leaves a track in the detectors if it is long-lived, or generates a hard lepton when it decays.
Finally, if the lightest neutralino is the MSSM-LSP and decays into the LSP in the detectors, the signature can be distinguished from the backgrounds by searching for events with multiple (displaced) photons.
All of these possibilities and future sensitivities on them will be studied elsewhere.


\section{Conclusion}
\label{sec:conclusion}

The SUSY not only solves the hierarchy problem between the Planck and electroweak scales, but also explain the 3--4$\sigma$ deviation of the muon $g-2$. 
The latter suggests that superparticles have masses of $\Order(100)\GeV$, whereas the Higgs boson mass of $126\GeV$ and the LHC results on the direct SUSY searches indicate that they exist in $\Order(1\text{--}10)\TeV$.
In this letter, we examined the parameter regions where the muon $g-2$ is explained with squarks decoupled. 
The muon $g-2$ anomaly is solved when electroweak superparticles are light and $\tan\beta = \Order(10)$. 
They are searched for by the LHC J- and L-searches. 
We found that the wino is constrained to be heavier than $240\text{--}260\GeV$ by the J-search under the approximate GUT relation on the gaugino masses, and than $150\text{--}200\GeV$ by the L-search irrespectively of the gluino mass. 
In addition, if sleptons are produced by decays of charginos and neutralinos, the L-search provides a severe constraint, and the bound can be as tight as $M_2 > 600\GeV$.
Consequently, the muon $g-2$ parameter space starts to be constrained by the LHC independently of details of the SUSY models. 

We also discussed future prospects on the collider searches for the superparticles which are relevant for the muon $g-2$.
A wide parameter region is expected to be examined at the 13--$14\TeV$ LHC.
In particular, the SM boson channels of the electroweak superparticle decays such as the $Wh$ productions are promising for the study. 
However, the searches are challenging when the Higgsinos are degenerate with the lightest neutralino, or if both of the wino and the Higgsino are heavy.
We need further studies to explore the whole parameter space where the muon $g-2$ anomaly is solved. 

The study in this letter is motivated by the muon $g-2$ anomaly. 
The $\gtrsim 3\sigma$ deviation may originate in the uncertainties.
Currently they are dominated by those from the measurement and from theoretical calculation of the hadronic contributions.
The former will be improved by the experiments at Fermilab~\cite{LeeRoberts:2011zz} and J-PARC~\cite{Iinuma:2011zz}.
The reduction of the uncertainty in the hadronic vacuum polarization contribution requires more experimental data for the cross section of $e^+e^- \to {\rm hadrons}$, e.g., at VEPP-2000~\cite{Shatunov:2000zc}.
The evaluations of the hadronic light-by-light contribution have been improving (see Ref.~\cite{deRafael:2012cg} for a recent review), and lattice calculations are in development~\cite{g-2_lattice}.
If the anomaly is confirmed in future, the SUSY is one of the most attractive candidates for the solution.
This letter is a first step towards the LHC searches for the scenario,
and we hope it is useful for further studies.

\section*{Acknowledgment}
This work was supported by JSPS KAKENHI Grant 
No.~23740172 (M.E.),
No.~21740164 (K.H.), No.~22244021 (K.H.)
and No.~22--8132 (S.I.).
The work of T.Y. was supported by an Advanced Leading Graduate Course for Photon Science grant.


\providecommand{\href}[2]{#2}\begingroup\raggedright\endgroup

\end{document}